# Solar-bound weakly interacting massive particles: a no-frills phenomenology.

Juan I. Collar

CERN, EP Division
Geneve 23, CH-1211
Switzerland
E-mail: Collar@mail.cern.ch

and

Groupe de Physique des Solides
Université Denis Diderot (Paris VII)
2 Place Jussieu, Tour 23
Paris 75251, France

The case for a stable population of solar-bound Earth-crossing Weakly Interacting Massive Particles (WIMPs) is reviewed. A practical general expression for their speed distribution in the laboratory frame is derived under basic assumptions. If such a population exists -even with a conservative phase-space density-, the next generation of large-mass, low-threshold underground bolometers should bring about a sizable enhancement in WIMP sensitivity. Finally, a characteristic yearly modulation in their recoil signal, arising from the ellipticity of the Earth's orbit, is presented.

There is a mounting observational evidence, at all cosmological scales, for a large (90-99%) missing-mass component in our universe. Weakly Interacting Massive Particles (WIMPs) are one of the prime candidates for this Dark Matter (DM), specifically at the galactic scale [1]. Most experimental WIMP searches aim at the detection of the energy deposited in ultralow-background detectors by WIMP-nucleus elastic recoils, a process having a very low expected rate, of order < 10 recoils / kg detector mass / day for candidates not yet excluded. The default WIMP population searched for has an isotropic distribution in the galactic frame, forming part of an extended galactic halo, with local density $\rho_{halo} \sim 0.3$ GeV / $c^2$ cm$^3$ and a characteristic Maxwellian speed distribution with laboratory-frame dispersion velocity ~ 300 km/s. This population is expected to give rise to a recoil signal extending up to energies $E_{rec}$ ~ few tens to few hundreds of keV (depending on the WIMP and target nucleus masses), yet maximal at low energies, close to the detector threshold, which is typically in the few keV range of deposited energy, $E_{dep}$. This broad and rather featureless recoil distribution resembles the ubiquitous natural radioactivity background, a fact that largely reduces the experimental sensitivity.

The recoiling nucleus carries an energy proportional to the square of the velocity of the incident WIMP; an alternative hypothetical population of WIMPs with a lower dispersion velocity ~ 30 km / s and a similar local density would have the interesting effect of concentrating *all* of the expected counting rate into the first few keV of the energy spectrum. This effect is no different, from a practical point of view, from that brought forth by a large improvement in detector resolution when searching for a discrete-energy faint signal (e.g., as in going from scintillator to semiconductor spectroscopy). It is nevertheless evident that such a signal would lie well below the threshold energy $E_{thr}$ of most present and planned large-mass

WIMP detectors[1], especially in those sensitive only to the fraction of recoil energy going into ionization, which in the case of conventional germanium detectors is just $E_{dep} \sim (1/6) E_{rec}$ for $E_{rec} \sim 12$ keV, and even smaller for dedicated scintillators [3]. Fortunately this does not apply to all cases: planned bolometers based on superheated superconducting grains (SSG) [4], when operated at temperatures < 1 K have in principle the ability to readily detect recoil energies in the few tens of eV range; this ability has been ascertained down to $E_{rec} \sim 1$ keV using monochromatic neutron irradiations at 40 mK [5]. Similarly, detectors based on superheated droplets [6] and sapphire bolometers [7] have sub-keV thresholds that allow for the detection of this putative signal.

Provided such a low-$E_{thr}$ device, a reduction in WIMP dispersion velocity of one order of magnitude would result in an increase of the low-energy "signal", i.e., the *differential* rate ($dR / dE_{rec}$, expressed in recoils / keV / kg of detector mass / day) of up to two orders of magnitude. This interesting possibility, first contemplated by Griest [8], might facilitate the exploration of part of the WIMP parameter space predicted by supersymmetric extensions of the standard model [1] even without further improvement on current levels of background or resorting to background rejection techniques. It is risky to extrapolate the observed background in existing large-mass detectors (the "noise") to this as-of-yet unexplored region below $E_{rec} \sim 10$ keV, but in the case of ultralow-background germanium detectors, no unaccountable sudden rise near to threshold is observed or expected after the electronic and microphonic noise are taken into consideration [9]. Particularly, partial energy deposition (via Compton scattering) by high-energy photons or cosmogenic tritium contamination contribute negligibly to this spectral region [9,10], the only evident spectral feature being a smooth rise below $E_{dep} \sim 40$ keV.

---

[1] For instance, the CDMS detectors [2] feature a present effective $E_{thr} \sim 30$ keV, while the DAMA scintillators [3] have $E_{thr} \sim 10$ keV.

This is compatible with elastic scattering by the $\sim 10^{-5}$ neutrons / cm$^2$/ s from natural radioactivity in rock walls and neutron-producing muon interactions in the detector shielding. If this is indeed the dominant low-energy background in present WIMP detectors, an increase in it of no more than ~1.5 is expected in going from $E_{rec}$ = 10 keV to $E_{rec}$ = 1 keV. In other words, to anticipate a "noise" of ~ 0.5 counts / keV / kg / day at $E_{rec}$ ~ 1 keV seems realistic [7]. It seems justifiable then to assert, for the sake of argument, that given the existence of the aforementioned WIMP population (at a local density $\sim \rho_{halo}$), low-threshold bolometers will enjoy the advantage of an increased "signal-to-noise" ratio by up to two orders of magnitude.

In a seminal paper, Steigman *et al* [11] studied the dynamic behavior of heavy neutrinos accompanying the gravitational collapse of the protosolar nebula. Since the escape velocity at the Earth from the Sun's gravitational potential is ~42 km/s, if heavy neutrinos (or by extension a generic WIMP) were trapped during the formation of the solar system and managed to survive until the present epoch, they would make up an eminently interesting objective for low-threshold WIMP detectors. The most effective trapping mechanism put forward in [11] is the dissipationless change in the statistics of particle orbits naturally produced by a rapidly changing gravitational field; in this scenario the conditions for the capture of a WIMP with velocity v during the collapse of a region (of ordinary matter plus DM) of size R are:

$$v^2 < -(\partial \phi / \partial t) R / v; \qquad v < R / t_f. \qquad (1)$$

where $\phi$ is the gravitational potential and $t_f$ is the free-fall time scale. In other words, the WIMP must not leave the scene during the collapse and must have a velocity less than the escape velocity. This implies, in the framework of the conventional understanding of solar system formation, that all DM within 0.1 pc moving initially slower than ~0.3 km/s would be efficiently trapped and concentrated into bound orbits [12]. The final conclusion in [11] was that this trapped population would lead to a large local density enhancement with respect to free-streaming halo WIMPs:

$$\rho_{SB} / \rho_{halo} \approx 0.3 \cdot 10^{-3} \, \eta^{3/2} \left[ R / r_E \right]^{3/2} , \qquad (2)$$

where $\eta = t_f / t_c \sim 1/5$ is the ratio between free-fall time and the duration of the collapse. For a protosolar nebula of size $R \sim 10^{17}$ cm and an Earth orbit radius $r_E \sim 1.5 \cdot 10^{13}$ cm, the predicted local density of solar-bound WIMPs, $\rho_{SB}$, would then be ~ 15 times larger than $\rho_{halo}$.

In contrast to this result, in a posterior analysis by Griest [8] a general argument favoring $\rho_{SB} \ll \rho_{halo}$ was given, based on Liouville's theorem and the assumption of slow collapse. Griest's own attempt at rederivation of Eq. (2) yields:

$$\rho_{SB} / \rho_{halo} \approx 10^{-3} \frac{(1-f)^{3/4}}{4} \left[ R / r_E \right]^{3/4} \approx 0.2, \qquad (3)$$

where $f \sim 0.1$ is the initial ratio of DM to total matter. It must be born in mind that besides all the assumptions and approximations made to arrive to these estimates, large uncertainties exist in the fundamental parameters of the pre-solar nebula (its initial mass, R, $t_c$ and f) [13].

Subsequently, Gould *et al* [14] recognized the complexity of determining the exact value of $\rho_{SB}$, identifying four sources, three forms of internal evolution and two sinks for this solar-bound WIMP population, the sources being a) evaporation from the Earth via collision with nuclei, b) orbit capture from the galactic halo, c) three-body capture from the halo and d) WIMPs captured during solar system formation. To these one might add the occasional solar-system or proto-solar nebula crossing by low-speed, high-density WIMP aggregations [15] which might enhance $\rho_{SB}$ in a clumpy halo scenario [16]. The internal evolution would be determined by e) scattering with nuclei in the Earth, f) close gravitational interaction with the Earth and, g) long-range gravitational interaction with the planets. The two sinks are h) capture via scattering in the Earth and Sun, leading to possible posterior annihilation in its core, a detectable process [1], and i) three-body expulsion. Perhaps the most important practical conclusion [17] from all the above is that due to purely gravitational diffusion by encounters with the Earth, Jupiter and Venus, solar-bound WIMPs with velocities relative to Earth 12 km / s < u < 30 km / s are expected to have a local phase-space density equal to that of free-space (halo) WIMPs of the same velocities, i.e., a negligibly small ~0.05% of $\rho_{halo}$. For $u < (2^{1/2} - 1)v_E \sim$ 12 km/s diffusion into unbound orbits is kinematically impossible, whereas for u > $v_E \sim$ 30 km/s (the mean orbital velocity of the Earth) the diffusive time scales are longer than the lifetime of the Earth, $\tau_E \sim$ 4.6 Gyr. Hence, solar-bound WIMPs outside the small range of velocities 12-30 km/s may have preserved their primordial $\rho_{SB}$ at the time of solar system formation, whatever it might be. Finally, it is important to remark that Jupiter would be very efficient at "cleaning" (via scattering) the inner solar system of bound WIMPs with orbits reaching out to it, the evaporation time for this process being of only $T_{evap} \sim 10^{-3} \tau_E$ [12]. All planets inner to Jupiter, including the Earth, have $T_{evap}$ > 7.5 $\tau_E$ [12].

Most recently, Damour and Krauss [18] have paid special attention to the sub-population of WIMPs that undergo grazing collisions with the outer surface of the Sun, losing enough energy to fall into Earth-crossing orbits, followed by planetary gravitational perturbations so that their orbits can no longer cross the Sun. This leads to a long term survival greater than $\tau_E$ for orbits inner to Jupiter's.

Unfortunately, the estimated density for this particular family is seemingly $\rho_{SB}/\rho_{halo} < 0.1$ for WIMPs not yet excluded by the most sensitive underground searches[2].

This amalgam of information must be translated into something more practical for the DM experimentalist, if advantage is to be taken of the possible increase in signal-to-noise at low $E_{rec}$. Of particular interest would be a compact expression for p(u)du, the velocity distribution at Earth of surviving solar-bound WIMPs, which would enable us to calculate the differential rate of interaction in WIMP detectors for any arbitrary value of $\rho_{SB}$. As it turns out, the necessary information is at hand; any WIMP in this population must obey, at a minimum, the following conditions:

i)     $r_{aph} < r_{Jup}$ (the WIMP aphelion must not reach Jupiter, to avoid evaporation during $\tau_E$ [12])

ii)    $r_{peri} > r_S$ (the WIMP perihelion must be larger than the Sun's radius, otherwise scattering over $\tau_E$ may lead to accretion and posterior annihilation in the solar core [1,19])

iii)   $r_{aph} > r_E$ and $r_{peri} < r_E$, i.e., the orbits must be Earth-crossing, to be of practical interest.

For Keplerian orbits the perihelion (distance of closest approach to the Sun) and aphelion (apex) are related to the WIMP orbital invariants E and J by:

$$r_{aph,peri} = \left(\frac{J^2}{GM_S}\right) \Big/ \left(1 +/- \sqrt{1 + 2\frac{J^2}{GM_S}\frac{E}{GM_S}}\right) \quad \text{(top sign for perihelion)} \quad (4)$$

---

[2] WIMP candidates leading to larger values of $\rho_{SB}$ are already excluded by the most stringent WIMP limits [3], which are more restrictive than those used in [18].

(E is the WIMP energy in the gravitational field divided by WIMP mass, J is its angular momentum divided by WIMP mass, $M_S$ is the solar mass and G is the gravitational constant). Expressing E in units of $GM_S/r_S$ and $J^2$ in units of $GM_S r_S$, Eq. (4) takes the compact linear form (common for both aphelion and perihelion):

$$E = \left(\frac{1}{2x^2}\right) J^2 - \frac{1}{x}, \qquad (5)$$

where x is the adimensional distance $r_{aph,peri}/r_S$. It is now straightforward to formulate the three minimal orbital conditions listed above in this convenient E, $J^2$ parameter space (with $r_E \approx 216.6\, r_S$ and $r_{Jup} \approx 1126.6\, r_S$), as is done in Fig.1; they restrict the allowed orbits into a *closed* region in parameter space. This invites to numerically sample this small region homogeneously, obtaining for any point in phase-space the WIMP speed during Earth-crossing in the Sun's reference frame, $\omega$, via the relation $E = (\omega^2/2) - (GM_S/r_E)$, to then transform this velocity to the laboratory (Earth's) frame by means of $\vec{u} = \vec{\omega} - \vec{v}_E$. In this last step, the angle $\theta$ between the Sun-Earth pointing vector, $\vec{r}_E$, and $\vec{\omega}$ is given by $J^2 = \omega^2 r_E^2 \sin^2\theta$ and the azimuthal angle of $\vec{\omega}$ around $\vec{r}_E$ is assumed to be homogeneously distributed (i.e., WIMP orbits are not restricted to the Earth's orbital plane). By repeating this sampling procedure it is possible to build the desired p(u)du of Fig.2, which is optimally described by the expression:

$$p(u)du \propto e^{-\left|\frac{u+w_1}{\beta u + w_2}\right|^\alpha} \qquad (6)$$

with α=2.8, β= - 0.17 and $w_1$= - 45.17 km/s, $w_2$= 27.02 km/s. This compact formula differs from the numerically computed p(u)du by less than 5% for all u > 10 km/s (Fig. 2).

The ellipticity of the Earth's orbit affects the distribution in Eq. (6) with a yearly periodicity: first of all, the lower boundary of the allowed region in E, $J^2$ space (Fig.1) depends on $r_E$, which undergoes a yearly ±1.7 % oscillation around 1 AU. Second, the value of $v_E$ oscillates between ~ 30.3 km/s (on ~January 3rd, the time of the Earth's perihelion) and 29.3 km / s (on July 3rd, the aphelion). Finally, the angle between $\vec{v}_E$ and $\vec{r}_E$ undergoes a very small yearly change of order 1%. All these changes translate into a fluctuation of the available phase space by ±0.6 % and of <u> around 41.34 km/s by ±1.3 % (upper signs are for January, lower for July), and can be effectively taken into account (Fig.2) by expressing the speeds $w_1$, $w_2$ as time-dependent functions:

$$w_1 = a_1 + b_1 \cos[\psi(t-3)]$$
$$w_2 = a_2 + b_2 \cos[\psi(t-3)], \quad (7)$$

where $a_1$= 45.197 km/s, $b_1$= 0.614, $a_2$= 27.080 km/s, $b_2$= 0.374, $\psi = 2\pi / 365.24$ radian / day, and t is the number of days after January 1st. The percent change in WIMP interaction rate induced by this modulation is shown in Fig.3 for several target materials constituent of planned low-threshold detectors. These results are representative of any WIMP interacting via scalar couplings (e.g., a Dirac neutrino or a neutralino with a Z-ino-Higgsino mixture) and are not influenced by the magnitude of the coupling.

A second geographically-dependent diurnal modulation in <u>, of maximal magnitude O(1%) (not treated here but straightforward to calculate), should arise from the partial daily alignment and counteralignment with $\vec{v}_E$ of a laboratory's rotational velocity around the Earth's axis (~0.45 km/s near the equator). While this diurnal modulation in the interaction rates is smaller than its yearly counterpart even for an optimally located equatorial laboratory, its daily periodicity can amply compensate for this by providing a rapidly growing statistical significance in relatively short runs[3]. As an added advantage, no directional sensitivity in the detector is required for its identification.

The assumptions made to obtain Eq. (6) (homogeneity of the distribution of orbits within the shaded region in Fig.1 and of azimuthal angles) are necessary evils: leaving aside the uncertainties in the initial proto-solar nebula conditions, a computer simulation of the exact p(u)du, including the sources, evolution and sinks discussed above, is per se a challenge to present-day systems [14]. Equations (6) and (7) must then be looked upon as first-order approximations to the gross features of p(u)du, encapsulating first-principle information common to all possible solar-bound populations, and hopefully useful for the experimentalist in extracting limits on $\rho_{SB}$ or searching for a tell-tale modulation signature. It is nevertheless important to estimate the extent to which a particular trapping and evolution mechanism can change a population's spectrum of energy deposition in a detector. Fig.4 displays these departures for the case in which p(u)du vanishes for 12 km / s < u < 30 km / s [17] and for a more compactly-packed population of orbits with an aphelion contained within 1/2 $r_{Jup}$ rather than $r_{Jup}$. It can be seen that neither provokes a dramatic change in $dR / dE_{rec}$, especially when a reasonable detector resolution is folded in. Features such as the maximum energy deposition or the spectral region affected by the modulation(s) remain relatively insensitive to the fine details of p(u)du, and would facilitate the identification of the mass of the responsible WIMP.

---

[3] The volume of data collected by several large-mass experiments is individually approaching the figure of 100 kg-y and therefore sensitivities of O(0.1%) to diurnal variations in WIMP interaction rate are expected in the near future [20]. The modulation produced by the rotational speed of the laboratory around the Earth's axis may then soon become a useful Dark Matter signature even for the conventional galactic halo population with dispersion speed ~300 km/s.

A final remark is in order: the present experimental limits on an spherically symmetric distribution of solar-bound DM (of any kind, including baryonic DM) arise from precision measurements of the motion of the exterior planets [12]. The bounds so obtained can be expressed as $\rho_{SB} < 1.4 \cdot 10^7 \, \rho_{halo}$ for DM interior to Neptune or $\rho_{SB} < 2.2 \cdot 10^7 \, \rho_{halo}$ for orbits within that of Uranus. These limits should be rapidly and largely improved for non-baryonic candidates by the first generation of low-threshold WIMP detectors.

**Acknowledgments:**

I would like to thank Vigdor Teplitz for useful discussions during the early stages of this work, which was partly supported by grant ERB4001GT965187 of the EU TMR programme.

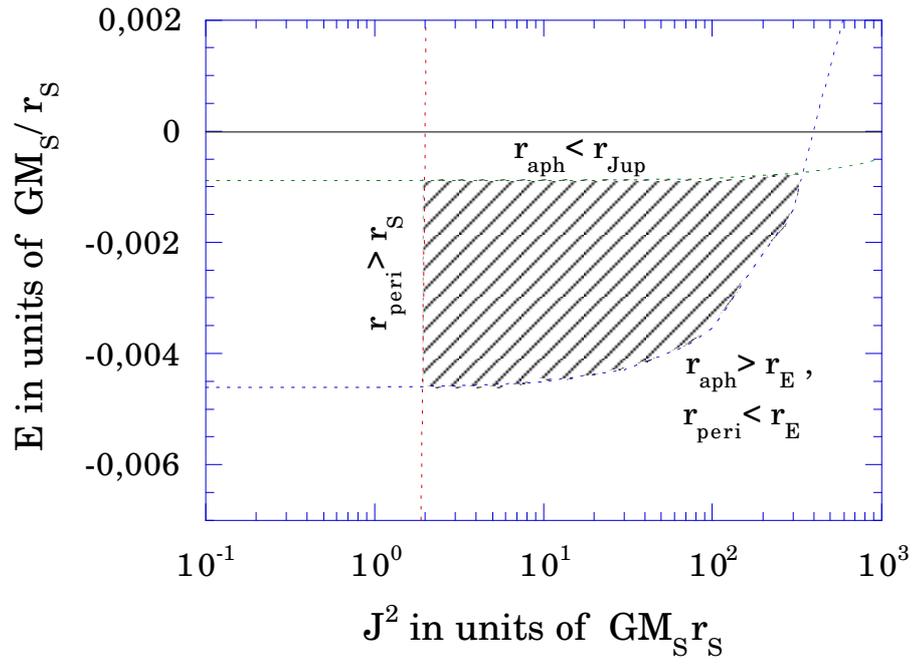

Fig.1: Available phase space for the orbital invariants of energy and angular momentum belonging to a detectable solar-bound WIMP population, under the minimal assumptions described in the text.

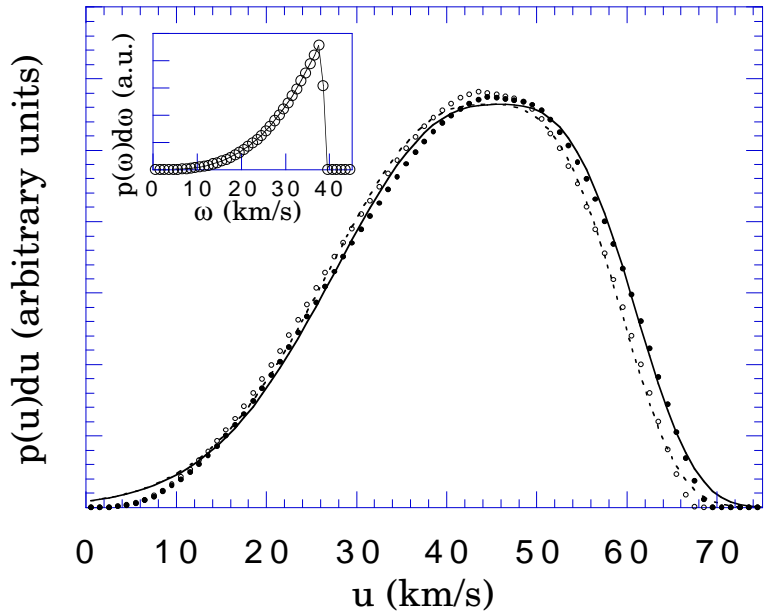

Fig.2: Speed probability distribution at Earth for WIMPs uniformly distributed in the hatched region of parameter space in Fig.1. Black circles correspond to January 3rd (the time of Earth's perihelion) and white circles to July 3rd (aphelion). The lines (solid = Jan., dotted =Jul.) are generated by the approximation to p(u)du described in Eqs. (6) and (7). The deviation of this fit is larger than 5% only for u < 10 km/s; WIMPs with such low speeds are not expected to leave a recoil signal above any realistic detector threshold. Insert: same distribution, but in the Solar reference frame.

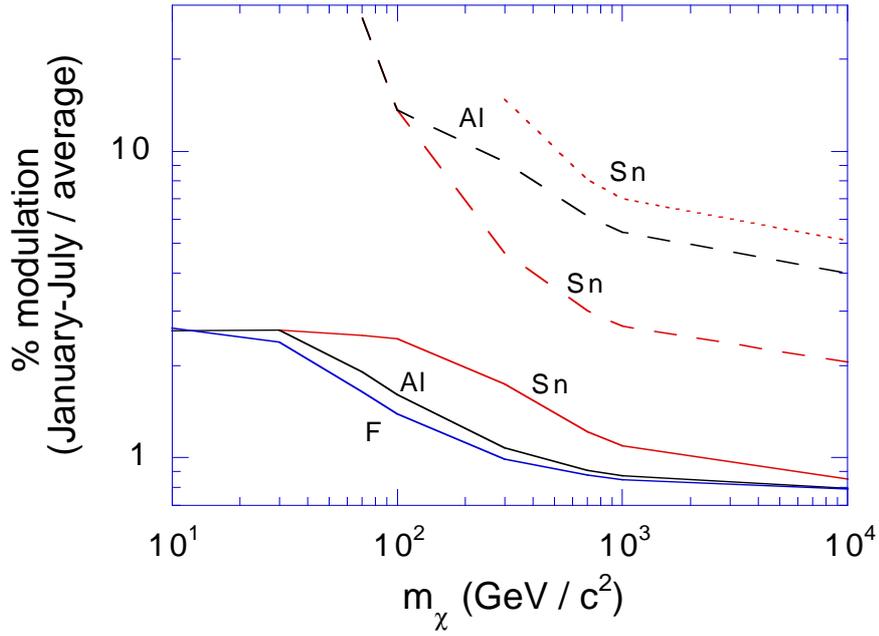

Fig.3: Magnitude of the yearly modulation in $dR/dE_{rec}$ arising from variations in p(u)du due to the ellipticity of the Earth's orbit, for several target elements to be employed by next-generation low-threshold devices. Solid lines correspond to changes in total detection rates for scalar-coupling WIMPs, dashed are for a detector threshold $E_{thr} = 1$ keV and dotted for $E_{thr} = 3$ keV. As in all WIMP modulations, the seasonal changes are maximal near the endpoint of $dR/dE_{rec}$ (albeit this is where the magnitude of $dR/dE_{rec}$ is minimal, making the identification of the modulation harder). No line is plotted when the endpoint in $dR/dE_{rec}$ for light elements is below the assumed $E_{thr}$.

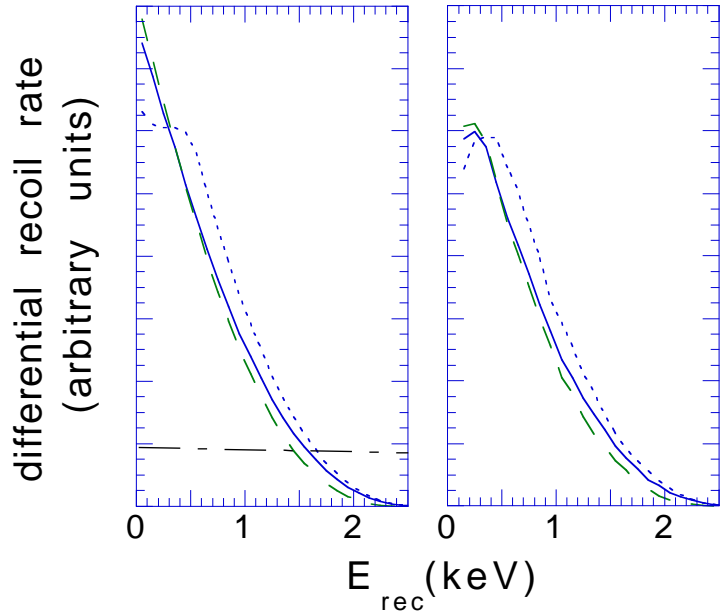

Fig.4: Left: $dR/dE_{rec}$ for a Sn target in the presence of a solar-bound population of WIMPs of mass 100 GeV/c$^2$ with predominantly scalar-couplings, for a fixed arbitrary $\rho_{SB}$. Solid lines are for the p(u)du given by Eq. (6), dotted lines are for the same but with no contribution from 12 km/s < u < 30 km/s, and dashed lines are for the case when WIMP orbits are contained within 1/2 the radius of the orbit of Jupiter, rather than reaching out to it. The dashed-dot line is the expected $dR/dE_{rec}$ from the conventional galactic-halo unbound population at $\rho_{halo} = \rho_{SB}$ and is offered as a reference for the increase in $dR/dE_{rec}$ brought about by solar binding (this increase is larger for heavier WIMPs). Right: Idem after folding in a conservative detector resolution of 0.3 keV FWHM.